\documentstyle[sprocl]{article}

\def\be{\begin{equation}}
\def\ee{\end{equation}}
\def\bea{\begin{eqnarray}}
\def\eea{\end{eqnarray}}

\begin{document}

\title{WHAT DOES D-WAVE SYMMETRY TELL US ABOUT THE PAIRING MECHANISM?
}

\author{K. LEVIN, D. Z. LIU and JIRI MALY}

\address{James Franck Institute, University of Chicago, Chicago, IL 60637,
USA}


\maketitle\abstracts{
In this paper we argue that d-wave symmetry is 
a general consequence of superconductivity driven by repulsive 
interactions.  Van Hove (or flat band) effects, deriving from the two 
dimensionality of the $CuO_2$ plane are important in stabilizing 
this state.  
By extending the original Kohn-Luttinger picture to a 2 D lattice, we 
find that 
the screened Coulomb term has important wave vector structure which 
leads to $d_{x^2-y^2}$ superconductivity
}

There is a growing body of  evidence to support the claim that  the 
pairing symmetry in the cuprates is most probably d-wave
wave\cite{Scalapino}, although some inconsistencies remain. 
These observations lead naturally 
to the question:  what constraints does this information provide on 
the detailed nature of the pairing mechanism? 
In this paper we investigate this important question based on
the Kohn Luttinger picture\cite {Kohn}
in which it was
argued that the
gap equation
\begin{equation}
\Delta_{\bf k } = 
-\sum_{\bf k' }V ({\bf k-k'})\left( \Delta_{\bf k'}\frac{\tanh\beta 
E({\bf
k'}) }{2E({\bf k'})} \right)
\end{equation}
could be satisfied for repulsive interactions $V > 0$ , provided $V$ 
contained some significant variation with wave-vector.  A reasonable 
choice for $V({\bf q})$ is the screened Coulomb interaction
$V_{eff}({\bf q}, \omega )= V_o({\bf q})/ \epsilon({\bf q},\omega)$
which in coordinate space
exhibits the well known Friedel oscillations; thus there are regions 
of negative  sign and superconductivity can take advantage of 
this attraction by forming anisotropic Cooper pair states. 

The magnetic pairing scenario\cite{Scalapino}, which has been widely 
discussed for 
the cuprates, may be viewed as a simple extension  of the 
Kohn Luttinger picture where the repulsive interaction (in wave 
vector space) is chosen to be the magnetic susceptibility
with peaks around the ($\pi/a$ , 
$\pi/a$) 
point.
As was shown by several 
different groups, 
this particular form for the wave-vector dependence in $V({\bf q})$ 
leads to a $d_{x^2-y^2}$  pairing symmetry. 

While this scenario evidently explains the  pairing symmetry, it is 
not at all clear whether it is the appropriate mechanism for the
superconductivity in  the  
cuprates.  Our 
group\cite{Radtke}  has extensively investigated this question based on 
the 
constraints imposed on the magnetism by neutron data\cite{Aeppli} in the copper 
oxides, and have inferred
that the magnetism appears too 
weak to be responsible for the high superconducting transition 
temperatures found in 
the cuprates.  

\begin{figure}
\vspace{2mm}
 \vbox to 5cm {\vss\hbox to 5cm
 {\hss\
   {\includegraphics{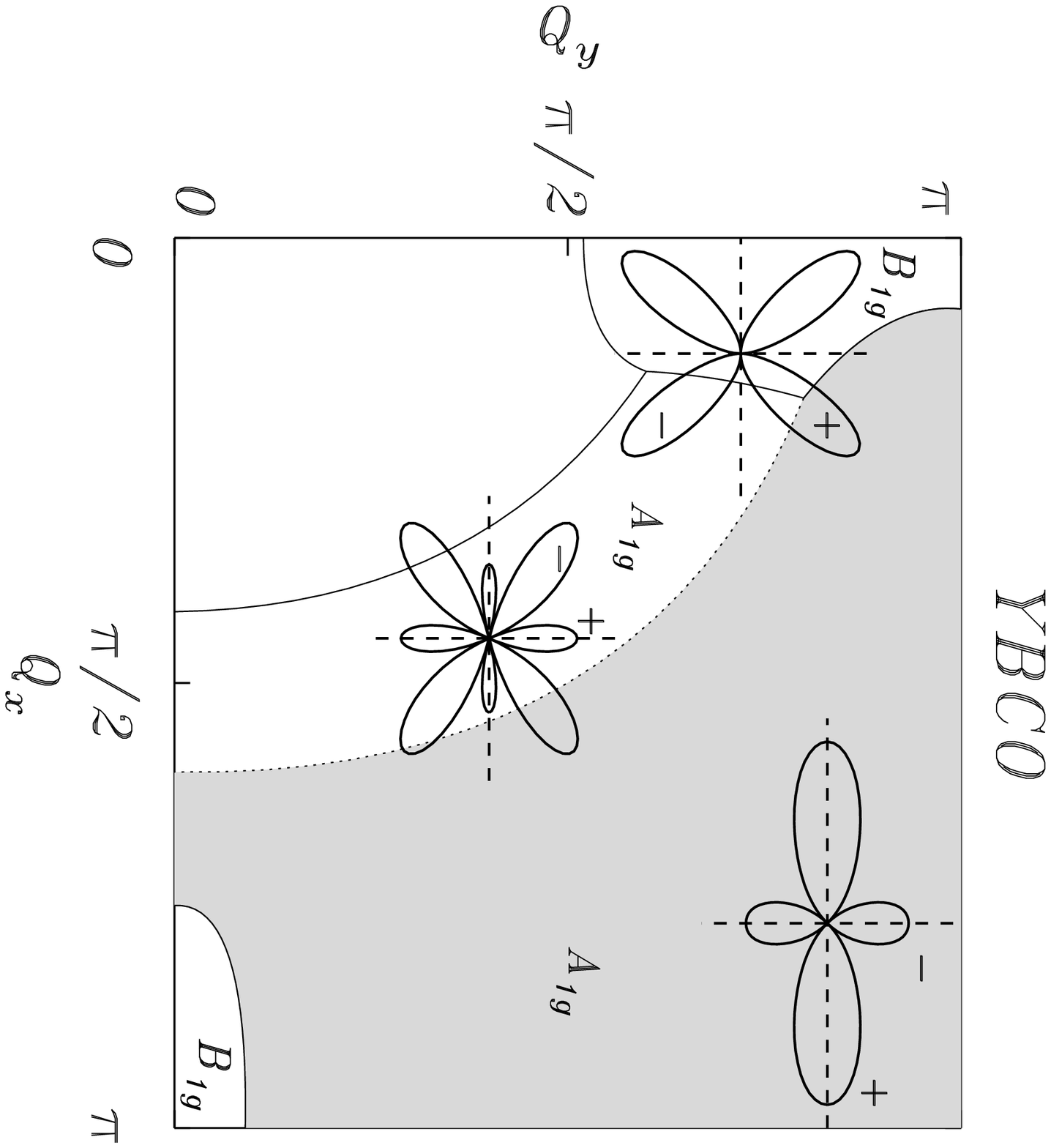}
   }
  \hss}
 }
\caption{
Phase diagram indicating gap symmetry for YBCO-like bandstructure.
\label{hs-f1}}
\end{figure}

Where,  then, does the $d_{x^2-y^2}$  symmetry originate?  To 
address this question we first build intuition by analyzing a simple 
``generic'' model which 
contains variable wave vector structure.  This wave vector structure 
is crucial for anisotropic superconductivity. Consider $V(k_{x}, 
k_{y})$   
given by
\begin{equation}
   V(k_x,k_y)= \frac{\lambda}{[1-J_o(\cos
(k_x\pm Q_x)+\cos 
(k_y\pm Q_y))]^2} 
\end{equation}
which is taken to be repulsive ( $\lambda > 0$ ) and to have maxima 
at 
variable  $Q_x $, $Q_ y$ with peak widths characterized by $J_ o$. 
The magnetic pairing scenario is a special case of this more general 
interaction with $Q_x $ = $Q_ y$  = $\pi /a $. As expected from the 
Kohn Luttinger picture, if this interaction is 
substituted into Eq(1), a variety of anisotropic pairing states will 
arise.  The various allowed symmetries\cite{Maly} 
are plotted in Figure \ref{hs-f1} as a function of $Q_x $, $Q_ y$.  Here the 
electronic 
structure is taken to be that of YBCO, near optimal stoichometry 
where the Van Hove singularities ( or flat band regions)  are in close 
proximity to the Fermi surface.

An important conclusion of this study is that the largest fraction of 
phase space is associated with $d_{x^2-y^2}$  pairing.  Thus, one may 
conclude that  this symmetry should {\it not} be related to any 
particular 
wave vector structure. It appears to be a robust consequence of 
superconductivity driven by repulsive interactions.  What stabilizes 
this $d_{x^2-y^2}$  pairing state over other candidate states? There 
are two important effects: (1) it is a state with a minimal number of 
sign changes in the gap function, ( as compared to the eight lobe s- 
state  ( $A_{1g}$) also shown) and consequently yields higher 
$T_c$'s for most slowly varying interactions
and (2) it is a state which takes maximal advantage of the high 
density of states associated with the flat band regions or "Van Hove"  
effects. Moreover, these flat bands are observed experimentally in a 
variety of cuprates\cite {photoemission}. 
It should be stressed that our 
early work\cite{old Physica C review} has noted that correlation 
effects contribute in an important way to pin the flat bands  near the 
Fermi energy, so that these effects should not be interpreted as 
simple (i.e.,  one electron)  Van Hove singularities. 

Despite these suggestive results, a more microscopic picture is clearly 
desirable.  The most natural source for a  repulsive interaction which 
may drive the superconductivity in the cuprates,  is the direct 
Coulomb interaction. 
Moreover, it is 
essential to our work that we begin with the full long range Coulomb 
interaction, rather than the Hubbard model approximation. We have, 
therefore, generalized the Kohn Luttinger calculation based on the 
screened Coulomb interaction to the case of a two dimensional, tight 
binding lattice. 
Here we find two important effects play a role 
in our analysis: local field and Van Hove contributions. Local field 
effects lead to a matrix form for the dielectric constant\cite{Sham}, 
so that higher  
Brillouin zone contributions or Umklapp processes are included.

\begin{figure}
\vspace{2mm}
 \vbox to 6cm {\vss\hbox to 6cm
 {\hss\
   {\includegraphics{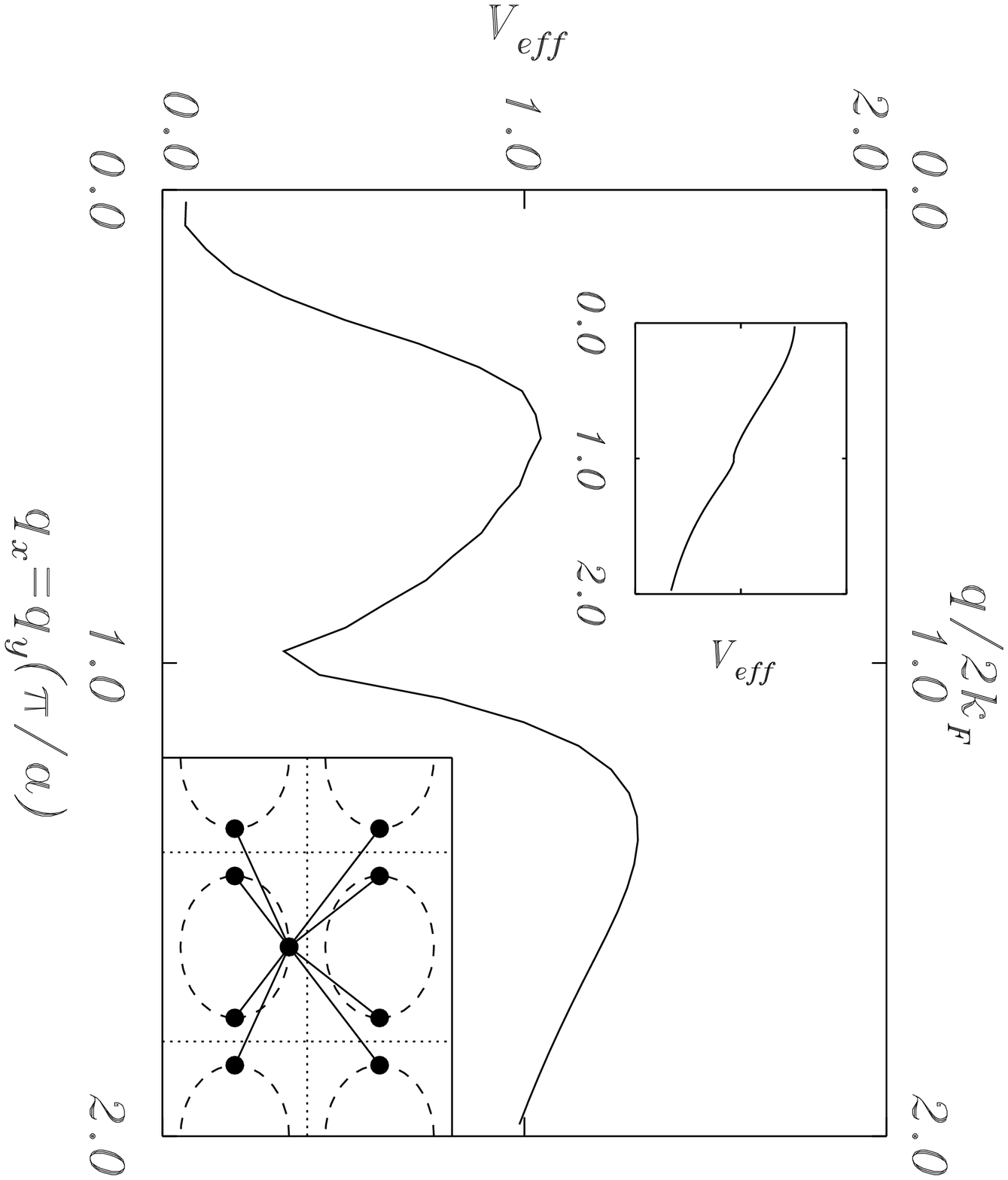}
   }
   {\includegraphics{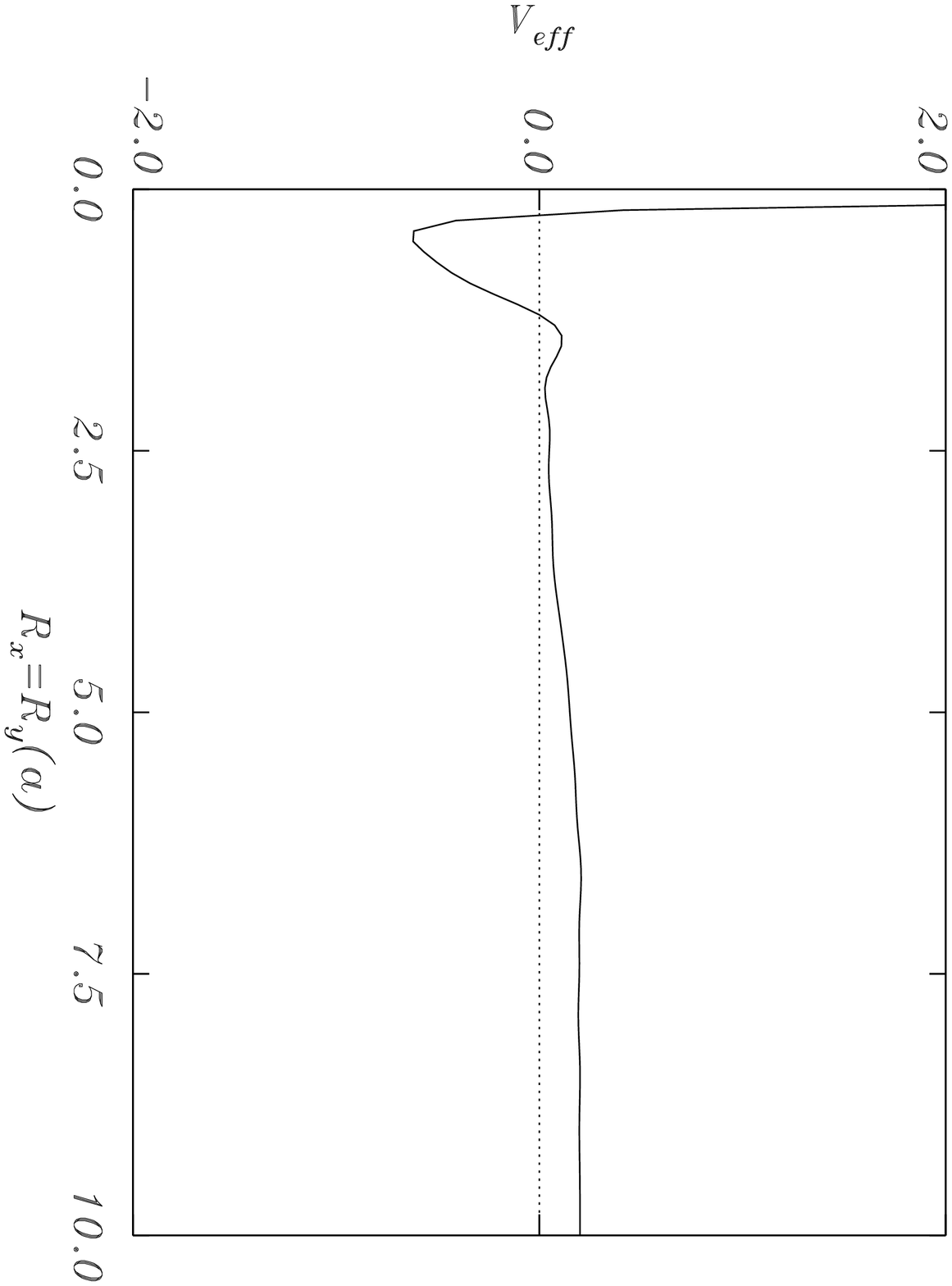}
   }
  \hss}
 }
\caption{(a) Screened Coulomb interaction for LSCO model bandstructure. 
Upper left inset: $V_{eff}$ for 3D
jellium. Lower right inset: demonstrates how Van Hove effects
are enhanced by Umklapp
processes. 
(b) Screened Coulomb interaction in real space in the static limit.
\label{hs-f2}}
\end{figure}

In Figure 2a  we plot our results for the screened Coulomb 
interaction for LaSrCuO (at optimal stoichiometry) and, as a point of comparison, 
for three dimensional jellium. The results for the YBaCuO family are found to be similar. The effect of the flat band regions 
enters not only in the density of states effects discussed in the 
context of the simple generic model in Figure 1, but also directly 
into the 
interaction itself.  These large density of states regions cause a dip in 
the 
interaction associated with enhanced screening. This dip is then 
followed by a maximum, slightly above the ( $\pi/a$ , $\pi/a$ ) 
point. 
Moreover 
these Van Hove effects are enhanced further through higher zone or 
Umklapp processes as shown in the inset. Finally, and most 
importantly, the superconducting instability associated with this 
wave vector structure is indeed, the $d_{x^2-y^2}$  state. 

In Figure 2b we plot the screened Coulomb interaction in co-ordinate 
space in the static limit. This figure shows the expected Friedel 
oscillations, which (in the Kohn Luttinger language) 
may be viewed as driving the superconductivity. The characteristic 
energy scale beyond which the Friedel oscillations become 
significantly altered, so that the d-wave pairing is no longer stable, is 
of the order of 4t (where t is the nearest neighbor hopping). This cut-
off 
thus represents a sizeable fraction of the plasma frequency.

What about the characteristic size of $T_c$?  
We are in the process 
of 
introducing 
corrections to the mean field equation which include phase fluctuation effects and which will enable us to compute $T_c$ more reliably.
These phase fluctuation effects are manifested in "pseudo-gap" behavior
away from optimal stoichiometry. It is only when such a 
scheme is in hand that we can with some confidence make further 
progress on the pairing mechanism.

In summary, this paper has demonstrated the generality of 
the $d_{x^2-y^2}$  pairing symmetry and has suggested an alternate 
route to d-wave pairing via the long range Coulomb interaction.  
Whether this is all or only a component of the pairing is too soon to 
say, but it is clear that direct Coulombic effects will act in concert 
with any other underlying d-wave pairing mechanism. 

This work is 
supported by the National Science Foundation (DMR 91-20000) 
through the
Science and Technology Center for Superconductivity.


\end{document}